# "I am both here and there" Parallel Control of Multiple Robotic Avatars by Disabled Workers in a Café

GIULIA BARBARESCHI* and MIDORI KAWAGUCHI*, Keio University, Graduate School of Media Design, Japan

HIROKI KATO, Ory Laboratory, Japan

MASATO NAGAHIRO, OriHime Pilots, Japan

KAZUAKI TAKEUCHI, Ory Laboratory, Japan

YOSHIFUMI SHIIBA, Ory Laboratory, Japan

SHUNICHI KASAHARA, Sony Computer Science Laboratories, Inc., Japan

KAI KUNZE, Keio University, Graduate School of Media Design, Japan

KOUTA MINAMIZAWA, Keio University, Graduate School of Media Design, Japan

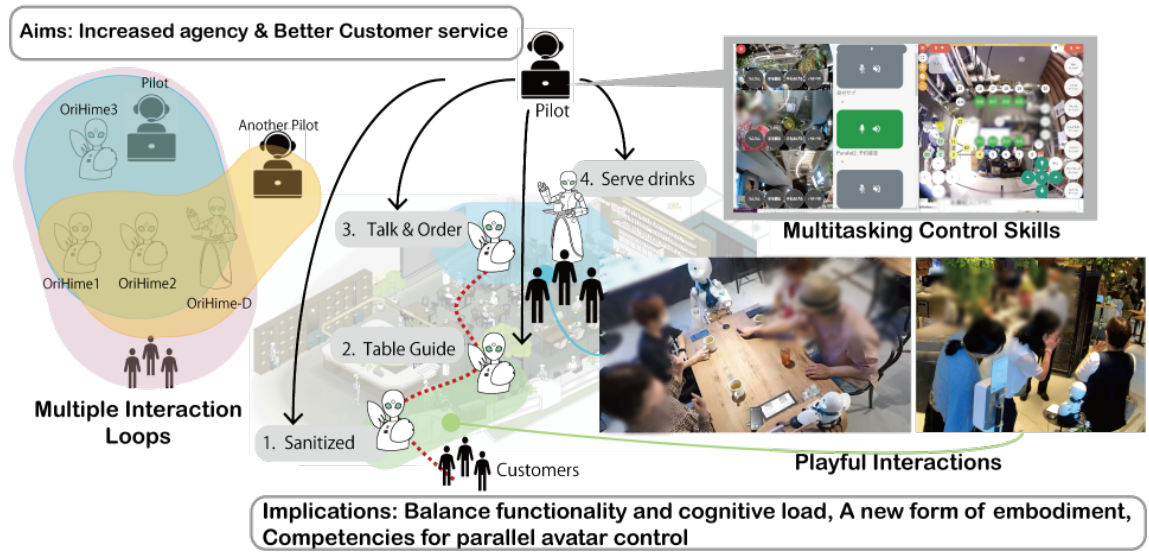

Fig. 1. A visual abstract for this paper. We co-developed a parallel avatar system that enabled disabled workers to control multiple avatar robots in a café

---

*Both authors contributed equally to this research.

Authors' addresses: Giulia Barbareschi, barbareschi@kmd.keio.ac.jp; Midori Kawaguchi, midori@kmd.keio.ac.jp, Keio University, Graduate School of Media Design, Yokohama, Japan; Hiroki Kato, Ory Laboratory, Tokyo, Japan; Masato Nagahiro, OriHime Pilots, Tokyo, Japan; Kazuaki Takeuchi, Ory Laboratory, Tokyo, Japan; Yoshifumi Shiiba, Ory Laboratory, Tokyo, Japan; Shunichi Kasahara, Sony Computer Science Laboratories, Inc., Tokyo, Japan; Kai Kunze, Keio University, Graduate School of Media Design, Tokyo, Japan; Kouta Minamizawa, Keio University, Graduate School of Media Design, Tokyo, Japan.

Robotic avatars can help disabled people extend their reach in interacting with the world. Technological advances make it possible for individuals to embody multiple avatars simultaneously. However, existing studies have been limited to laboratory conditions and did not involve disabled participants. In this paper, we present a real-world implementation of a parallel control system allowing disabled workers in a café to embody multiple robotic avatars at the same time to carry out different tasks. Our data corpus comprises semi-structured interviews with workers, customer surveys, and videos of café operations. Results indicate that the system increases workers' agency, enabling them to better manage customer journeys. Parallel embodiment and transitions between avatars create multiple interaction loops where the links between disabled workers and customers remain consistent, but the intermediary avatar changes. Based on our observations, we theorize that disabled individuals possess specific competencies that increase their ability to manage multiple avatar bodies.

## 1 INTRODUCTION

The pivotal "*Cyborg Manifesto*" by Haraway 1991[31], introduces the figure of the cyborg as "*cybernetic organism, a hybrid of machine and organism, a creature of social reality as well as a creature of fiction*"(p.149). Several scholars, Including Haraway herself, have argued that, as a result of the interconnected relationships we have with technology in most aspects of our lives, we are all cyborgs [5, 9, 31, 38, 48]. Critical Disability scholars and activists have explained how disabled people have a longer history of integrating technology into their lives and their bodies, compared to most non-disabled individuals [43, 64, 97]. In fact, many assistive technologies are seen as an extension of the body that become integral parts of the embodied experience of the individual [2, 66, 73, 75, 76]. Disabled people's embodied cyborg experiences are often associated with the use of assistive devices, which are directly connected or in proximity of one's body such as wheelchairs, prosthetic devices, and cochlear implants [4, 6, 19, 45, 62]. However, previous research has shown that embodied connections also exist between disabled individuals and their digital or physical avatar bodies including simple tele-presence devices and more complex robotic agents[68–70, 79, 88].

Researchers have developed different types of mobile tele-presence devices and robotic avatars with the purpose of extending the capabilities of people with disabilities and enabling participation in activities that would otherwise be inaccessible[26, 59, 90, 102]. However, most of the existing research on disabled people's use of tele-presence devices and avatar robots, is limited to usability studies involving novice users, rather than disabled individuals who have extensive experiences controlling their own robotic avatars [11, 26, 90, 92, 102]. This severely limits our ability to understand the specific competencies that disabled people develop to better control their "remote bodies" and, to examine the impact of embodied experiences linking the disabled individual and their robotic avatar[70, 92, 96].

Recent technological advances expanded the possibility of connection between users and physical or digital robotic agents beyond linear systems where one body can only inhabit one avatar[83, 85]. More specifically, these novel systems can allow one individual to move between different avatar bodies (re-embodiment), multiple individuals to share the control of a single avatar body (co-embodiment), or even having one individual controlling multiple avatar bodies simultaneously (parallel embodiment)[49, 83].

These new control paradigms could offer significant advantages to disabled users, allowing them to consciously negotiate between benefits and limitations of different bodies, physical or digital, depending on situational needs [70]. Despite this, studies exploring the use of different forms of embodiment for robotic avatars have thus far rarely included disabled participants, and have primarily been limited to laboratory-based experiments not representative of real-world scenarios [29, 51, 53, 83].

In this paper, we present the results of an innovative study which tested the use of a system allowing disabled workers to control multiple robotic avatars inside a café'. The system was purposefully developed with the support of disabled workers to increase their agency during regular operations, expanding their ability to carry out multiple actions in sequence. The "*pilots*" leveraged the multiple robotic avatars to accompany customers through their regular journey from entering the establishment, being assigned a table, ordering drinks, and being served. To unpack these interactions, we leverage data including interviews with disabled pilots, surveys from the café's customers, and video of the café operations collected during the experimental launch of the parallel system. Our results show how pilots' transitions between avatars and the simultaneous control of multiple bodies create unique interaction loops where the connection between people (pilots and customers) always remains consistent, but the avatar agent constantly changes. We also highlight how the skills and specific competencies of expert disabled pilots are essential to the successful management of parallel operations, despite the increased cognitive load associated with multi-body-multi-tasking. Finally, we observe how parallel interactions do not only support the optimization of customer service operations, but also create unique moments of surprise and wonder, for both pilots and customers, which are often delightful, but occasionally confusing or frustrating.

Based on these results, we extract broader design implications for the development of new embodied cyborg experiences that leverage more-than-linear control paradigms. We theorize that, through their everyday experiences, disabled individuals have developed specific competences that offer significant advantages in the management of non-linear avatar interactions.

This paper makes the following contributions:

- The first co-development and implementation of a parallel system that allows disabled users to control multiple robotic avatars simultaneously in the context of a real world café
- The first exploration of the interactions between disabled pilots, their robotic avatars, and in-person café's customers
- Reflections and design implications for the development of future avatar systems to expand embodiment possibilities for disabled individuals

## 2 RELATED WORK

This work builds on three different areas of research: (Dis)abling experiences with robotic avatars and telepresence devices, understanding different forms of avatar embodiment, and Disability and telework. Throughout this paper, we interchangeably use identity-first and person-first language in acknowledgement of the different preferences of individuals and disability communities[20].

## 2.1 (Dis)abling experiences with robotic avatars and telepresence devices

In 1980 Minsky described the concept of *Telepresence* as giving a "*remote participant a feeling of actually being present at a different location*"[50]. Robotic avatars represent the natural evolution of this concept by providing a physical form to the surrogate persona of the user [81]. Over the years, the development of robotic avatar technology specifically aimed at supporting disabled users has attracted significant attention across the engineering, robotics and HCI communities[21, 30, 87, 89, 100]. Proposed applications for telepresence devices and robotic avatars have varied significantly including but not limited to remote school attendance[54], supporting communication between older adults with dementia and their families [52], social connectedness [26], conference attendance [70], remote work [84],and the performance of various physical tasks for self-assistance [67].

Although the specific domain of application might differ between studies, the advantage offered by these technologies is that they potentially enable disabled users to bypass some barriers that would otherwise prevent them from taking part in certain activities. In some cases, these barriers might be linked to the physical inaccessibility of a particular venue or location. For example, both Ng et al [56] and Friedman & Cabral [26], explored the use of telepresence devices and robotic avatars to increase the accessibility of museums and archaeological sites. A telepresence device can also enable the individual to re-negotiate capabilities between their physical and avatar bodies. Both the auto-ethnographic experience of Rode [70] and the interviews with children unable to attend school in person carried out by Ahumada-Newhart & Olson [1], show how access to a telepresence device or an avatar robot might allow individuals to choose an alternative form of social participation when health concerns or bodily affordances might make in-person attendance too challenging.

Although telepresence devices and robotic avatars might, in many situations, present a way to navigate access issues, they also present accessibility barriers of their own [1, 10, 70, 92, 98]. One of the most obvious difficulties is linked to the need for consistent and reliable connectivity. Loss of connection, from either side, can leave the disabled user cut off from their own avatar, which in turn will be stuck in their last location [92]. While for someone with a mobility impairment being blocked by a physical barrier is definitely a frustrating and disabling experience, being physically present allows the person to call attention to the issue and ask for help if necessary [80, 94]. However, the loss of connection between physical and avatar body removes one's ability to attract the attention of bystanders, building managers, or anyone who might be able to offer support. Suddenly the disabled person becomes completely invisible and their agency is completely removed [70]. Visibility is a familiar double edged issue for many disabled people [22]. On the one hand, disabled individuals are consistently overlooked due to ableist prejudice, at the same time, and for often the same reasons, they can also be subject of unwanted attention and curiosity [3, 13, 22, 36]. The experiences of disabled children using telepresence robots in the classroom largely resonated with this seemingly opposing dichotomy [55] . While the avatar robots enabled them to attend class and meet their friends, the children also stated that it was much easier for them to be overlooked in their avatar form, as their physical presence was not fully acknowledged [55]. Yet, the peculiar look of their telepresence robot caused them to attract curious looks and negative comments from other children, leading them to discontinue their use in favor of *"traditional"* services offered to students unable to attend in preson [55].

Like all forms of technologies that disabled people integrate into their daily lives, telepresence devices and robotic avatars, require certain skills to be used effectively [26, 47, 101]. As the embodied connection between the disabled user and the avatar robot increase, users develop new competencies and personal strategies that allow them to expand the interaction capabilities of their own avatars [2, 7, 65]. Rode [70] poignantly describes how, through years of use, she was able to perfect a *"whole-body robot knock"*, as her telepresence avatar robot did not have hands that could be used for that purpose.

Despite the fact that time will affect both users' skills and sense of embodiment, most studies investigating the use of telepresence devices and robotic avatars by disabled individuals involve novice users and are relatively short in duration [102]. This can limit our ability to understand how a growing sense of embodiment and connection between disabled users and their robotic avatars could open the door to new opportunities in a variety of domains.

### 2.2 Understanding different forms of avatar embodiment

When we use the term embodiment in relation to the use of robotics avatars, we refer to the ability of a user to feel present and perform actions in a different location from where their physical body exists, through the use of a robotic agent [23]. Most widespread forms of avatar embodiment involve one-to-one connections between a single user and a robotic avatar [49]. These forms of avatars embodiment have been used in an incredible variety of contexts, from classrooms and museums, to the bottom of the ocean, and even in outer space [1, 44, 46, 71].

Other forms of avatar embodiment feature multiple users simultaneously sharing a single robotic avatar, often described as co-embodiment [25]. The purpose of these systems is generally to support the collaboration between two or more individuals, enhance training experiences, and augment performance of users [28, 85]. However, previous research by Fribourg et al [25] highlighted how participants struggled to correctly estimate their sense of agency when sharing the control of an avatar with another individual or autonomous agent, and that discrepancies between the expected and resulting movement of the avatar body could lead to extremely confusing experiences. Interestingly, the study by Luria et al [49] also showed that even when individuals witness co-embodiment interactions of multiple agents in a single device without directly experiencing it, they perceived it as confusing, and occasionally unsettling due to the creation of unbalanced power dynamics.

One-to-many avatar embodiment can be supported in a sequential fashion, usually referred to as re-embodiment, or in a simultaneous manner, known as parallel embodiment [49, 83]. Systems that support avatar re-embodiment allow users to selectively migrate between a set of robotic avatars that might be placed on different locations or have different characteristics that can be leveraged to perform various tasks [14, 15, 60]. While the ability to move between multiple avatar robots at will offers definite advantages, previous studies have shown that it requires high levels of situational awareness, and it can easily result in overall decreased performance due to the complexity of the interaction[14].

Finally, parallel embodiment involves similar challenges to re-embodiment, but these tend to be magnified by the fact that the user is simultaneously trying to control two or more robotic avatars [27]. When multiple avatar bodies are used to perform the same action, the difficulties involved with parallel control are significantly reduced [51]. If parallel embodiment is used to perform different actions using multiple avatar bodies, mechanisms for reducing users' cognitive burden are usually introduced in the form of automatic movement assistance [83], supported integration of multiple control paradigms [53], or simplified control options that trigger pre-determined commands [27].

### 2.3 Disability and tele-work

The ongoing COVID-19 pandemic has made many of us familiar with the realities of remote work. However, many disabled people have been using technology to access the workplace remotely long before the start of the pandemic [41, 86]. Almost 30 years ago, Hesse 1996[34] stated that telework could represent a viable strategy to drastically increase the accessibility of many places of work. Although such a revolutionary vision never fully materialized, technology has been leveraged by many disabled people to access workplaces that would have otherwise been inaccessible[86].

Overall, telework encompasses a broad variety of technologies ranging from document editing software, to file sharing and different forms of mediated online communication, including telepresence [16, 70, 95, 104]. Recent studies by Zolyomi et al [104] and Das et al [17] investigated the experiences of neurodivergent professionals using videoconferencing software. Similarly, Vogler et al [93] and Rui Xia Ang et al [72] highlighted how current videoconferencing technologies fall short of supporting the needs of Deaf and hard of hearing workers in relation to clarity of communication, being able to attract attention non-verbally, and supporting the involvement of interpreter teams in collaborative tasks. Despite the emergence of high-quality research specifically focusing on how disabled workers use videoconferencing technology, little attention has been dedicated to mobile telepresence devices and robotic avatars.

In her ethnographic account Rode described the experience of using a telepresence device during an academic conference, as well as in her departments both in the US and the UK [70]. Her detailed account clearly articulates how the physical characteristics of the robotic avatar and the capabilities and limitations afforded to the user influence work interactions [70]. This corporeal dimension becomes particularly relevant as we consider the use of these technologies in the context of the service and hospitality industry, where tasks might require physical labor and in-person interactions with customers.

Prabakar & Kim [61] proposed the design of a 180cm tall wheeled Tele-bot with a humanoid appearance (featuring 52 Degrees of Freedom across more than 20 different joints including head, shoulders, arms, and waist), speakers, microphones and cameras to be used by veterans or law enforcement agents with mobility impairments for patrolling duties. The robotic avatar was developed and further refined by Sundarapandian et al 2016[82] , but never tested with disabled officers.

The only example of robotics avatars controlled by disabled workers in the service industry was proposed by Takeuchi et al [84] where authors presented the results of a two week trial involving 10 disabled participants who used mouse, touch screen or gaze input to control avatar robots acting as waiters in a café via their own computers. Results from surveys administered to disabled workers showed that the systems could provide a fulfilling opportunity for employment and keep the workload manageable for the individual. However, pilots could only control a single avatar robot as part of their shift which limited their ability to carry out customer service operations. Moreover, the study only focused on testing the feasibility of the system and did not provide extensive insights on customers and pilots experience, or investigate the capabilities of the pilots that allowed them to effectively control the avatars.

## 3 DEVELOPING A PARALLEL AVATAR SYSTEM FOR ACCESSIBLE CAFÉ TELEWORK

In this paper we describe the development and implementation of a parallel control system in a real world avatar café. In the following sections, we first describe the characteristics of the café, and the development of the new parallel system.

### 3.1 Café settings and regular operations

The Avatar Robot Cafe DAWN ver.$\beta$ officially opened its doors to customers in June 2021 in Tokyo and features both in-person workers and disabled tele-workers. The management of the café aims to showcase how technology can be harnessed to create inclusive employment opportunities even in a sector that traditionally requires the performance of physical tasks that might be inaccessible to many. The café's features six tables, seating up to 4 people each, that can be booked in advance through the website or via phone. Customers who book these tables are assisted by disabled teleworkers who control OriHime avatar robots (commercially available through https://orihime.orylab.com/) each in charge of different operations.

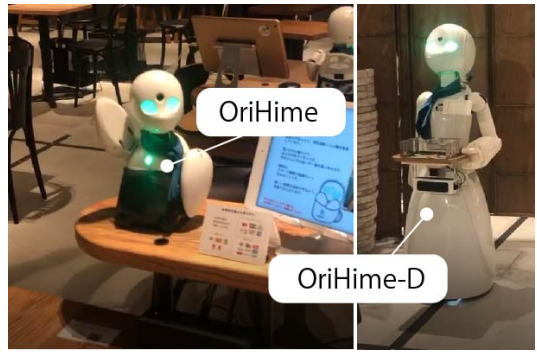

Fig. 2. A small OriHime Avatar placed on a table and a large OriHime-D used in the café

To date, the café employs 65 disabled individuals living in different parts of the country who leverage the avatar robot OriHime to work their shifts. Each shift lasts from 1 to 4 hours depending on the availability of the person and can feature up to 7 disabled teleworkers, commonly referred to as *pilots*, each in control of a different avatar robot. The café features two different types of OriHime avatar robots, which are used for different purposes.

Small OriHime avatar robots, such as the one showed on the left side in Figure 1, are placed on each of the six tables which can be reserved by customers. These avatar robots are approximately 23cm tall and 17cm wide. The neck of the robot has two Degrees of Freedom (DoF) allowing for head movement on the sagittal and frontal plane, the two arms have only one DoF and can be moved using a series of pre-programmed simple command such as "raise hand", "wave", or "flap both arms". These avatar robots are used by disabled pilots to explain the menu, collect the orders, and entertain the customers for the duration of their booking. One of each of these smaller OriHime are also placed respectively by the entrance of the café and before entering the seating area. The OriHime placed by the entrance is used to greet customers and ask them to sanitize their hands, whereas the OriHime placed just before the seating area is used to check customers' booking and assign them to a table.

Larger avatar robots, known as OriHime-D, are approximately 120cm tall and 50cm wide. They feature two motorized omniwheels at the base to enable them to move across the café floor. Similarly, to the smaller OriHime, the neck of the OriHime-D has two DoF, but this larger model also has two movable arms with 6 DoF each (2 at the shoulder, elbow, and wrist), which can be controlled using pre-prepared motions, namely "raise one hand", "bye-bye", 'hold up fists', "and "power pose". These larger avatar robots are used by pilots to carry food and drinks from the kitchen to the customer's table.

Like more traditional telepresence devices, both smaller OriHime and large OriHime-D are equipped with cameras, microphones and speakers. Disabled workers are able to access available avatar robots in the café through a dedicated web application. Once a pilot "logs-in" their assigned OriHime for their shift, they are able to access the view of the camera, enable/disable speakers and microphones and activate pre-programmed motions through onscreen buttons. For both small OriHime and large OriHime-D, pilots can control head movement through an onscreen D-pad. Pilots in charge of the OriHime-D can also drive them around the shop to move between tables by using the motorized wheels at the bottom.

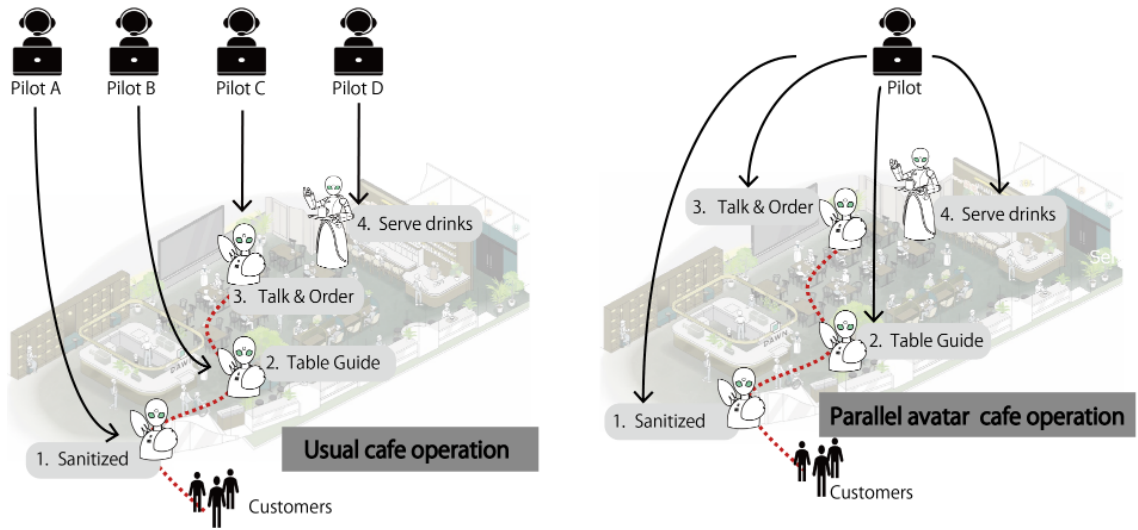

Fig. 3. Two diagrams showing both the regular (on the left) and parallel (on the right) flow of operations as customers move around the café and the members of staff in charge of different steps

During a regular shift, with 7 disabled teleworkers, 3 would be controlling as many small Table OriHimes, 2 would be controlling larger OriHime-Ds, and 2 would be in charge of the OriHime located by the entrance and outside the seating area. Once a group of customers enters the café, one of the pilots, controlling the small OriHime avatar robot placed by the entrance will greet them and ask them to sanitize their hands. The customers will then move towards a the small OriHime placed before the seating area, with the role of a floor manager, who will assign them to their table. Once at the table the disabled pilot operating the corresponding Table OriHime will explain the menu, take their orders, and notify the kitchen. Another pilot operating one of the OriHime-D will then bring the food and drink to their table. A simple diagram showing customer's journeys in the café and their interactions with staff is shown in Figure 3.

About six months after the opening of the café, some of the pilots became accustomed to the operation and found that the current system which limited their connection to a single OriHime did not allow them sufficient agency in managing customers. Pilots noted how, in other cafes, floor staff would be able to follow customers along their journey, and began playfully experimenting for new strategies with in-person staff after closing time. The management of the café purposefully allows pilots to freely play around with avatar robots after closing time to support the development of personal connections between members of staff, facilitate upskilling, and give staff the opportunity to propose new ideas. During one of these sessions one of the pilots experimented with simultaneous control of two OriHime and perceived both the entertainment and functional value of the experience, which led to the current study.

### 3.2 Development of a New System for Parallel Avatar Control

After the pilots became interested in the potential of parallel control the management of the café reached out to researchers at the Keio Graduate School of Media Design, and members of the Sony Computer Science Laboratories, Inc. in April 2022 to explore the technical feasibility of the idea.

Authors, MK, SK, KM had previous experiences developing computer systems that support the user's ability to control multiple avatars [51, 83] and proposed that a similar system could be successfully employed in the context of the café. Authors HK, KT, YS, who had significant experience managing café operations, including MN, who had been a café pilot for over 3 years, felt that a system that enabled workers to migrate between different OriHime avatar robots could be effective in managing customer journeys without altering the overall flow of operations, or adding new avatar robots with different features and capabilities.

The idea of developing and testing a parallel control system was then pitched to the other pilots to gain their feedback and identify a sub-set of volunteers who had an interest in supporting the development and testing of the system. In total, 6 pilots joined the effort to design and test the system. Four pilots were female and two were male, ages were between 19 and 53 years, and their experience working in the café ranged between 6 months and 3 years. Due to both medical reasons and existing access barriers, all participants had limited their ability to leave the house, which prevented them to be able to seek in-person employment. P1 and P2, accessed their computer using a combination of eye-gaze and adapted clickers. P3, P5 and P6 utilized standard mouse and keyboard interfaces, and P4 preferred to use a tablet's touchscreen. All pilots stated that before beginning to work at the café they had been struggling to find satisfactory employment opportunities as disabled workers. P2, P3, P4, and P6 had engaged in different forms of remote work, but disliked the fact that jobs available to them were extremely solitary and involved little chance of interaction with others. Unsurprisingly, all pilots stated that entertaining customers was one of their favourite parts of working at the café, alongside camaraderie with other members of staff both in-person and remote. Beyond their part-time job in the café, both P4 and P6 utilised the OriHime avatars for work in other hospitality establishments, whereas P1, P2, and P5 engaged in other forms of remote work.

The development process was carried out over a period of four months. The initial phase of the development focused on determining the characteristics of the system in relation to the workflow of the café'. The second phase was centered on the building the user interface, and the final stage focused on pilot training.

*3.2.1 Architecture of the system.* The parallel avatar control system enables pilots to create multiple peer-to-peer connections to all the robot OriHime and OriHime-D available in the café. As the pilot connects from their home, they can see the full list of the avatar robots currently turned on. The architecture of the system is shown in Figure 3.

During the initial phase of development P1, P2, and P5 participated in the discussions with members of the research team to decide on ways to break down various customer service tasks. Pilots felt that maintaining the existing flow of operation would facilitate learning and tested the practical feasibility of having up to four OriHime interfaces simultaneously open. As pilots felt that the task was manageable, the flow of operation was finalised as follows:

- The Entrance OriHime (1) greets customers and reminds them to sanitize their hands
- The Manager OriHime (2) checks their reservation and assigns them to a table
- The Table OriHime (3) collects customers' orders and entertains them
- The Serving OriHime-D (4) delivers drinks to the customers

As certain OriHime avatar robots (Entrance, Manager, and Serving indicated by number 1, 2, and 4 in Figure 3) need to be used by different pilots depending to the group of customers they are interacting with, the system allows an incoming pilot to notify the pilot who is currently embodying one of these three avatars to request control as needed. If the pilot currently controlling the avatar consents to it, the system automatically switches the connection to the upcoming pilot.

*3.2.2 User Interface.* All pilots participated in this phase of development and each of them took part in at least one of the three testing sessions which were scheduled between May and June 2022. During these sessions, pilots performed tasks managing multiple UIs, identified pain points, and discussed potential modifications with the development team. Pilots also discussed feedback with each other, tried their different suggested modifications to find agreement around an UI that could be accessible to all of them, regardless of the different interaction modalities they utilized (gaze, touch,mouse,switches). From the start, it was decided that all small OriHime (Entrance, Manager, and Table) would retain the same UI to maximize efficiency and reduce confusion. In the original UI of the small OriHime pilots had access to a D-pad to control head movements, buttons for muting/un-muting microphones and speakers, and a series of buttons each corresponding to a different pre-programmed movement ("Nod", "Wave hand", "Raise hand"," Flap arms"). Movement buttons were kept consistent during various design iterations as pilots did not feel that they would have benefited from the addition of further options. However, pilots pointed out that controlling head movements through a d-pad could be difficult to do when the size of each screen was reduced due to the need to fit four different UIs on one monitor. To address this challenge, we implemented a different interaction modality that enabled pilots to simply click on the camera view of one of the robot avatars OriHime and control its head movement by moving their mouse cursor in the desired direction of sight. Buttons to mute and un-mute speakers were also moved to the side of the camera view to avoid creating unnecessary blind spots.

OriHime-D featured a more complex UI. A complete map of the café overlays the camera view and allows pilots to select different locations. Once the pilot selects a location, the OriHime-D follows the lines placed on the floor from the start to the end point. Manual driving is also possible through a D-panel placed at the bottom right of the screen. Similarly, to the small OriHime, OriHime-D can perform a series of pre-determined motions as the pilot presses the corresponding button placed on the right hand of the screen. Finally, another D-pad enables pilots to control the the head movement of the avatar and dedicated buttons allow for muting and un-muting microphones and speakers. As the size of the UI for the OriHime-D did not significantly decrease with the use of the parallel system, pilots recommended that no changes should be made to it to avoid the need for unnecessary learning. Figure 4 shows all four UIs of the different avatar robots as used by one of the pilots.

*3.2.3 Operations training.* All pilots took part in the operations training between May and June 2022. Initially, pilots came up with the idea to prepare and follow scripts when using each of the four avatar robots to smoothly manage customers operations. However, as their confidence increased, pilots found their own idea too restrictive and decided to abandon the scripts and develop their own ways to manage operations guiding and entertaining customers according to their own unique style. The week prior to the implementation study, five practice sessions were carried out to ensure that pilots could smoothly manage, not only their interactions with customers, but also the overall flow of operations at the café. This included passing control of shared avatars as needed, and timing drinks ordering and delivery to each table to minimise potential delays.

## 4 PARALLEL AVATAR CONTROL SYSTEM IMPLEMENTATION STUDY

To fully test the feasibility of the parallel avatar control system we decided to run an implementation study that would directly match the characteristics of a complete 1 hour shift where all 6 available tables would be booked by different groups of customers. To allow for sufficient time for pilots to guide their group of customers from the café entrance to their table we staggered the entrance of each group with 5-minute intervals (i.e. the first group entered at the start of the hour, the second group entered at 5 minutes past the hour, the third group entered at 10 minutes past the hour…). This format was chosen as it is in-line with the current system that allows different groups of customers to book each of the 6 available tables at a particular time slot. In the following sections we provide details about study participants, data collection procedures, and data analysis methods.

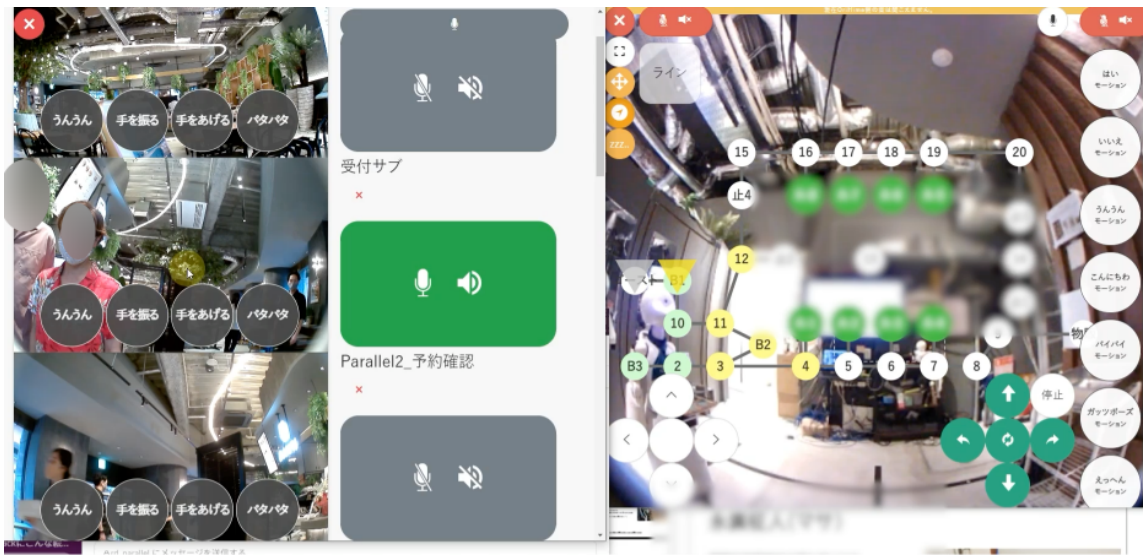

Fig. 4. Screenshot from one of the pilots showing the four UI of the corresponding OriHimes on the left side are displayed the Table, Entrance, and Manager OriHime, whereas the UI of the Serving OriHime-D is on the right side

### 4.1 Participants

In total 24 customers were recruited to take part in the study. Participating customers were recruited through a combination of café and project mailing lists, pilots' social media, announcements on universities and café's websites. Participants who responded to the adverts were screened to confirm that they were above the age of 18, able to provide informed consent and available on the day of the study. Participants meeting the inclusion criteria and who gave consent to take part in the study were issued an invitation to join one of the six tables and assigned a starting time for their entrance in the café. Most participating customers signed up to take part in the study as individuals, but 5 pairs of participants signed up together. Participants to joined the study in pairs were assigned to the same table, all other table allocations were randomly assigned. Four participants were seated at each of the 6 tables. Eight participants had never visited the café' before, whereas 16 were returning customers. Sixteen customers self-identified as female, 7 as male and 2 chose not to state their self-identified gender. Only 2 participants were aged below 35 years, 8 were between 36 and 45 years, 9 between 46 and 55 years, and 5 were between 56 and 65 years. Twenty-three participants were Japanese nationals. The remaining customer was non-native Japanese, but moderately fluent in Japanese. Eight participating customers stated that they had little to no interaction with someone who had a disability before the study, 18 customers had regular or frequent interactions with disabled people in their life, and 1 participant self-identified as a disabled person. Finally, 6 customers had never interacted with a robot or robotic avatar before the study, 9 had occasional interactions before, and 9 had frequent interactions with robots or robotic avatars previously.

### 4.2 Data collection methods

As we aimed to gather insights that captured the perspective of both pilots and customers as well as observe the interactions that occurred during the field study we employed a variety of data collection methods. Three 4K GoPro cameras were strategically and unobtrusively placed in different parts of the café to capture interaction between the customer's groups and the four different avatars controlled by the pilots. Video and audio data from pilots' own devices used to control the various avatars were also recorded using the OBS Studio Software to monitor their activity on the user interface and capture conversations they had with customers.

To gather feedback from customers and learn about their experiences during the implementation of the parallel control system, we asked them to fill out a short survey after the end of the study. The survey featured both open and closed questions including collection of demographic data, favorite moment during their café experience, suggestions for improvements, and differences with any previous visit in to the café (only for returning customers).

To understand pilots' perspectives, and to collect their recommendations regarding a permanent implementation of the parallel system as well suggestions for future developments, we conducted semi structured interviews in the week following the filed study. Interviews were conducted online via Zoom in Japanese, the native language of all the pilots. Author MK, a native Japanese speaker, led the interviews, with author GB, non-native Japanese with moderate proficiency in the language, also present during the interviews taking field notes and asking additional questions as needed. Interviews lasted between 55 and 70 minutes and were audio recorded. Questions during the interviews included pilots' overall experiences working in the café, explanations of how they normally interact with the UI of the OriHime avatar robots and how this was affected by the new system, challenges encountered when learning how to use the parallel control system and during the implementation study, overall impressions of the parallel control systems and their opinions about its future use in the café and beyond.

### 4.3 Data analysis

Data corpus consisted in video recordings from the three cameras placed in the café, video recordings of pilots' screens, audio recordings of pilots interviews, field notes taken by the authors, and surveys completed by customers. Videos from the café's camera were initially analyzed by author GB using interaction analysis to identify key moments in the exchanges between pilots and customers [42]. This method was chosen for its suitability to in-the-wild studies, use on video data, and for the purpose of understanding the nature of interactions between individuals and objects accounting for both verbal and non-verbal language [42]. Key moments identified through the café' cameras were also examined using videos from pilots' screen and inspected by author MK who produced full transcripts of the verbal exchange in Japanese and translate them to English. Quantitative data from customer surveys were analysed using descriptive statistics whereas qualitative data from open questions were translated from Japanese to English. Finally, recordings from interviews with pilots were transcribed verbatim by author MK and translated to English. All transcribed and translated data was analysed using deductive thematic analysis focusing on both semantic and latent interpretation [8]. Initial coding was carried out by author GB and initial codes where discussed and review with author MK, as themes began to emerge they were discussed with the other authors. Resulting themes from both the interaction analysis and thematic analysis were triangulated and discussed amongst all members of the research team until consensus on the interpretation was reached [24].

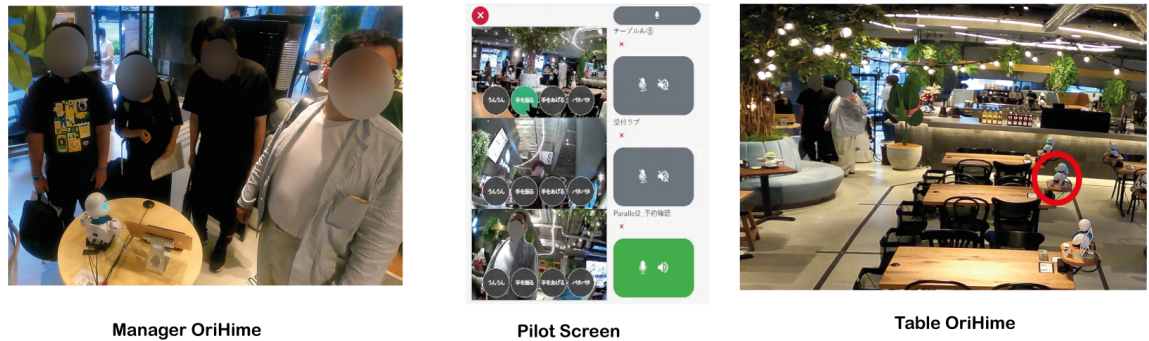

**Manager OriHime**  **Pilot Screen**  **Table OriHime**

Fig. 5. Picture showing the sequence of events and interactions as P2 helps the group of customers identify their assigned table in the café.

## 5 RESULTS

As a result of our analysis, we conceptualized three themes to help illustrate the interaction dynamics between pilots, their avatar robots and customers and their experiences during the implementation of the parallel avatar control system: "Layers of interaction between people and avatars", "Challenges, skills, and pride", "Purpose and play".

### 5.1 Layers of interaction between people and avatars

The architecture of the system allowed pilots to maintain contacts with customers at all time. However, while the connection between pilots and customers remained consistent, the presence of the avatar robots created unique interactions dynamics. One-to-one interactions where pilots and customers communicated through a single intermediary avatar, occurred at multiple stages, including when customers entered the café and were greeted by pilots controlling the Entrance OriHime, or during more prolonged conversations that took place at the table.

Different interaction dynamics were observed as pilots guided customers from the entrance, to the Manager OriHime's desk, to their assigned table. Pilots leveraged the features of the system to give verbal clues to customers through one avatar robot OriHime, while attracting their attention to another avatar robot OriHime using one of the simplified motions available. An example of this is illustrated by the sequence of events displayed in Figure 5. The second group of customers is standing in front of the Manager OriHime controlled by P2 (visible on the left side of the Figure, in the picture above the label Manager OriHime). As she explains to them which table they have been assigned, she says *"Your table, is table number 6. There is a OriHime on the table and she is weaving at you, can you see her?"* (P2). The customers turn to look towards the table area while P2 weaves at them through the corresponding Table OriHime. On P2 pilot screen, displayed at the centre of Figure 5, the user interface shows how P2 uses the active microphone and speakers of the Manager OriHime to talk to the group of customers, indicated by the green icons at the bottom, while using the button that gets the Table OriHime to raise its hand and weave, showed by the button in green at the top of the screen). One of the customers notices the avatar robot and says *"Yes, I can see it"* (G2 – C1) , while weaving back at the Table OriHime controlled by P2, highlighted by the red circle on the right side of Figure 5.

Other observed interactions involved more prolonged simultaneous parallel embodiment~~, as the pilots controlled two avatar robots~~. This occurred in particular as pilots were entertaining the customers through the Table OriHime, as they were also serving drinks through the Serving OriHime-D, as shown in Figure 6. These interactions involved much longer exchanges as pilots explained to customers the nature of the Parallel control system and customers curiously wondered about the sense of multiple presence created by the two avatars. The short dialogue except below illustrates one of these exchanges.

C1-G1: "You're moving there while you are talking here!" (pointing at both Table and Serving OriHime)

P1:" Yeah, right" (Switches on the microphone on both Table and Serving OriHime)

C2-G1: "Wow"

P1: "Now I'm operating both OriHime"   C1-G1: "Can you see me from both sides?"

P1: "Yes I am both here and there"

C4-G1: "It seems it really like many [name of P1] are here" (laughs)

C1-G1: "Yes he is doubled!"

P1: "Yes here is double me" (Laughs)

C3-G1: "It is really surprising to see [P1 name] from this side and that side, it means he see from both side and I am confused!!"

A more complex indirect form of interaction occurred among all pilots, customer groups and avatar robots at the broader café level. During a one-hour shift the Entrance and Manager OriHime, and the Serving OriHime-D were shared across all pilots and customer groups. This meant that at different stages of the shift, the same avatar robots were embodied by different pilots and used to interact with different groups of customers. Indirect interactions occurred between pilots as they needed to negotiate who was currently controlling a particular avatar robot, and manage their tasks accordingly. This turn-taking aspect was managed through a series of strategies, from adherence to pre-allocated timings for customers booking to notifications from the kitchen staff who let pilots know when drinks were placed on the tray of the OriHime-D ready to be served to customers. Figure 7 displays how these different interaction loops integrate both separate and overlapping aspects as the connection between people is unique, but the embodiment of avatar robots is shared by multiple pilots working in the café.

## 5.2 Challenges, skills, and pride

Although the parallel system boosted pilots' ability to manage customers' operations, being able to do so effectively required considerable skills. Despite initial worries, during training practice sessions pilots quickly realized that, while controlling multiple avatars simultaneously was not a trivial task, it built on competencies that they had already developed. Their existing competencies in controlling single avatar robot OriHime and their familiarity with café operations enabled them to rapidly built personalized strategies and gain confidence in their own abilities to manage customer journeys.

"When I saw the first explanation of the system and how we would use it, I wondered what would happen. It was hard to imagine for me. However, as I actually started working with it, I was able to understand what it was all about. I could see the advantages of being many alter ego instead of one. I was a bit confused at first, but in the end I was able to finish the event with the customers happily". (P4)

Despite one's skills parallel embodiment of multiple avatar robots involved a high cognitive load. P3 mentioned that one of the reasons that pilots in the café are given this title is linked to the fact that they need to manage a relatively complex control interface, similarly to an airplane pilot. However, their job also involved conversing with customers and entertain them, as it is often the case in the service industry.

"The good thing was that I was able to welcome and attend to customers from the entrance. The bad part was that I could not just concentrate on talking to customers. I had to keep glancing at the screen to see what was going on with the OriHime-D for serving the food." (P3)

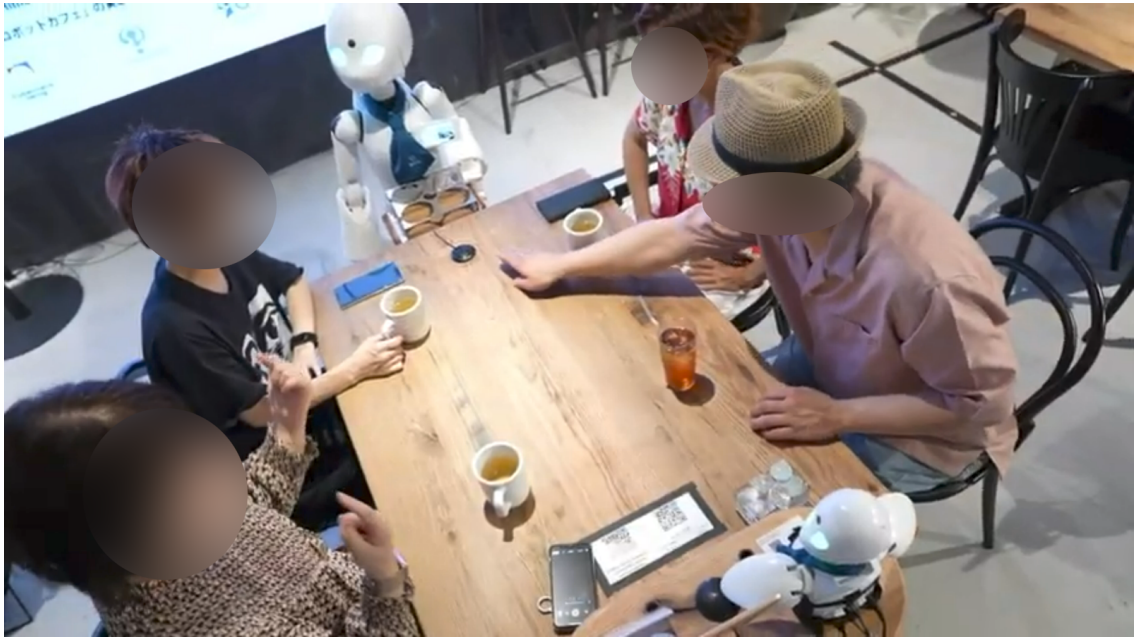

Fig. 6. Picture showing two OriHime avatars, controlled by the same pilot, facing each other and interacting with customers seated at the table

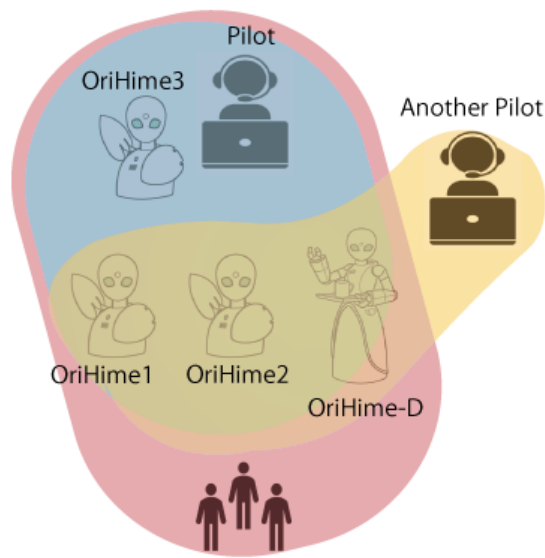

Fig. 7. Picture showing the interaction loops connecting pilots and customer groups through the exclusive use of Table OriHimes and the shared use of Entrance and Manager OriHimes, and Serving OriHime-D.

While challenges with managing different tasks through multiple avatars were somehow expected, pilots also pointed out how parallel embodiment also presented some difficulties that they did not anticipate. For example, P2 recalled that while she was driving the Serving OriHime-D from the table to the kitchen, a passing customer stopped to wave a greeting at her. Although she wanted to naturally respond to this interaction, she was unable to do so, as she was simultaneously talking to the group of customers at her table through the Table OriHime, and she was unsure how to explain this.

*"The most difficult part was when someone waved hello to me when I was carrying drinks with OriHime-D, but I couldn't talk because I was having a conversation in another OriHime at the same time. I can't stop the conversation unless I say, "Excuse me," and cut the conversation, but the other customer doesn't know that. My heart cried out, "Sorry, wait!""(P2)*

The complexity involved in parallel avatar operation was clearly perceived by customers. Several customers expressed amazement at how pilots were able to manage the process in such a smooth manner with one customers describing P6 as an "artist in the use of avatars" (G6-C1). In turn, pilot clearly felt proud of their own skills and devised unique ways to showcase them by making different avatars that they were simultaneously controlling interact with each other triggering surprised and admiring reactions.

*"One of the most memorable reactions of the customers for me was their confused but happy faces. They said, 'This OriHime is [P5 name] and this OriHime is also [P5 name]? That's amazing!' I was happy to hear that, and I was glad that I had worked so hard." (P5)*

### 5.3 Purpose and Play

The design of the parallel avatar systems was driven by the desire to expand pilots' agency and improve customers' experiences. Feedback gathered from the interviews with pilots showed that our first goal was largely achieved. While pilots acknowledged that certain tasks in the café, such as preparing drinks, were not yet accessible to them, the new system allowed them to take full charge of the floor operations. These increased responsibilities brought new challenges but also made their job more interesting and engaging, driving the desire for an even more flexible system that could further boost their capabilities.

*"My ability to serve customers has definitely increased. With a small OriHime, which is just left on the table, I need to stay the place all the time. In a normal café', I would have to go to the entrance to welcome visitors. Now I can do that kind of thing. It seems I can go to the exit when the customers leave. I think I felt closer to my alter ego. Now, I would like to make the café more active. I hope to become a master who customers call on for advice and talk to me about what they are doing and what fun they are having" (P3)*

Customers' responses also emphasized how the parallel system positively influenced their experiences by creating a sense of service continuity and enhancing their feeling of personal connections with the pilot. Some advantages were practical in nature such as reduced waiting time when being assigned a table, or ordering drinks. However, the biggest advantage they associated with the use of the parallel system was the impression of a more polished customer service experience, which increased the sense of intimacy.

*"The customer service itself felt very smooth. I felt a sense of security/reassurance when talking to the same person. There was a sense of hospitality when the same person was doing everything from the beginning to end and I felt I was in a luxury shop" (G4-C1)*

The parallel avatar system also created unique moments of surprise and amazement linked to the perception of multiple simultaneous presence. Like many novel experiences, parallel embodiment could generate a sense of confusion, which became more evident when, for example, customers quickly shifted their gaze between two OriHime avatar robots controlled by the same pilot, unsure about which one they should address. Pilots frequently described a sense of wonder, as having two avatars of themselves facing each other , made them feel like they were looking into a *"strange kind of mirror"* (P5).

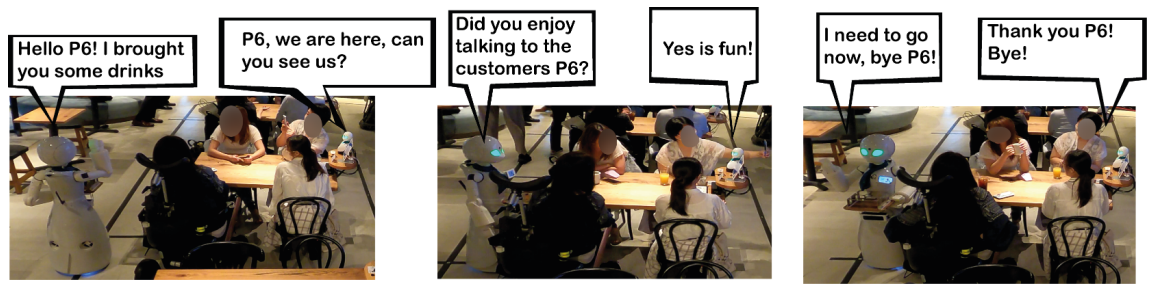

Fig. 8. Short sequence of pictures showing P6 enacting a playful dialogue between the Table OriHime and the Serving OriHime-D.

*"When I see my OriHime-D through the OriHime on the table it feels mysterious. I can't see myself in ordinary cafes. I look like a stranger, but I also look like me. I know it in my head, but I can't recognize it. I know I am myself when I move it and look at it."* (P6)

To share this sense of multiple presence with customers, pilots leveraged a variety of strategies. Projecting the voice of the pilot simultaneously from two OriHimes was an easy way to convey a sense of multiple presence. Returning customers who had interacted with one pilot on multiple occasions, were also able to detect similarity in the body language of different avatars controlled by the same individual (pilots also mentioned that they could spot these differences, when observing each other, explaining how this was akin to be able to recognise a friend from behind by the way they walk). Some of the more experienced pilots developed playful ways to entertain customers leveraging their multiple presence by alternating dialogue and movements between two avatars, mimicking a conversation between two different individuals. Figure 8 shows one of these playful interactions that P6 enacted to amuse the group of customers at her table.

## 6 DISCUSSION

Our work represents the first example of research involving expert workers with disabilities in the development and implementation of a parallel avatar system inside a real-world café. Our findings highlight the presence of key design implications for future systems involving the use of telepresence devices and robotic avatars featuring different control architectures especially in relation to understanding the embodied experiences of users, and balancing functionality with cognitive overload. Moreover, based on our observations we hypothesize that disabled individuals as a result of their daily practices interacting with technologies (assistive and non), possess specific competencies that enhance their abilities to leverage robotic avatars for a variety of purposes, including telework.

### 6.1 A different form of embodiment

A significant number of studies in HCI has explored the complexity of embodied experiences of disability and their technological implications [36, 37, 76, 78]. Research in HCI and beyond has also shown that embodied experiences of disability don't end at the skin, rather they organically incorporate assistive technologies including wheelchairs, prosthetic limbs, and mobile phones [2, 6, 12, 74, 77]. The study by Ringland 2019 [68] also documented how embodied experiences that involve digital avatars of autistic children who engage in the Autcraft virtual world can be considered as real, and in some cases more real, than the ones that take place in the physical world. The insights gained from pilots who took part in our study are broadly aligned with these trends. Pilots specifically referred to the avatar robot OriHime as their alter-ego, explaining how parallel interactions where two avatars were facing each other felt like a *"strange kind of mirror"*. Quotes presented above highlight how, when pilots looked at the OriHime avatar robots, they saw themselves. This indicates an extremely strong sense of embodiment and connection with their robotic alter egos. In many ways, this well matches findings from previous literature on disability and technology embodiment. However, there are some important differences that should be noted.

Several researchers have highlighted how personalization of assistive technologies, from cochlear implants to prosthetic arms, plays a significant role in transferring one's embodied identity to a physical device [32, 62]. Recent research by Zhang et al [103]. has shown that very similar practices exist in the personalization of virtual avatars, including the desire to incorporate assistive technologies, with the aims of better representing one's personality. Yet, all avatar robot OriHime and OriHime-D in the café look identical to each other, regardless of which pilot is currently embodying them. One distinctive aspect that emerged is that, although all OriHime avatar robots might look the same at first glance, both pilots and returning customers are able to recognize who is currently embodying them based on their body language.

Body language represents an extremely powerful and complex form of communication. Robotic agents can also leverage expressive motions to convey a variety of information to people interacting with them [91]. What emerged from the findings of our study is that pilots are able to express their unique body languages through their avatars using solely a combination of head movements and small sub-set of pre-programmed movements available through OriHime and OriHime-D. This potentially suggests that when developing robotic avatars there is not necessarily the need, at least from an expressive point of view, to incorporate extremely refined and fully unconstrained movements that might be extremely hard to control, especially using alternative computer interfaces. A skilled pilot can develop a sufficiently strong sense of embodiment to be able to convey information in their own unique way, with a relatively simple control interface.

With such a strong sense of embodiment between pilots and their alter ego, it is somehow surprising to notice that participants expressed no concern of discomfort about sharing the avatar robots with others. As shown in Figure 7, while Table OriHimes were controlled by one single pilot for the whole duration of the shift, Entrance and Manager OriHime, and Serving OriHime-D, were alternatively embodied by different pilots. Moreover, during multiple shifts in a single day, is commonplace for multiple pilots to also embody different Table OriHime. In a way, this is facilitated by the distance created by the liminal space that separates the physical world where the pilot interacts with their computer to control the avatar robot, and the physical world where the avatar robot moves [68]. Nonetheless, we argue that the lack of distinctive features between different OriHime and the transitional nature of the interaction between pilots and avatars plays a strong role in this dynamic.

Barbareschi & Inakage [2] explored how the connection between artists and their wheelchairs persisted even without direct interaction. In contrast, as pilots in our study moved between the multiple avatar robots, their embodied connection only existed with the avatar they were controlling. This strong but self-contained form of embodiment might open up new possibilities for systems that allow multiple users to share robotic avatars depending on their situational needs while still feel a strong sense of embodiment when they are actively using them. Currently, the parallel control system only enables pilots to connect with the OriHime avatar robots that are present in the café. If avatar robots with similar characteristics could be located in different environments allowing individuals to freely move between them as they become available, a larger scale system could be leveraged to reduce geographical and physical barriers in the ways we interact with the world.

### 6.2 Balancing the need for more functionality with cognitive demand

When we designed the parallel avatar system for use in the Avatar Robot Cafe DAWN ver.$\beta$ we were mindful of both the constraints inherent to the features of OriHime and OriHime-D avatar robots, and the ones created by the context of the café. Future researchers seeking to develop new parallel systems might face less constraints and have significantly more freedom to create not just a novel control architecture and user interfaces, but completely new types of robotic avatars with different features and capabilities. While situational constraints might be specific to the settings and the application domain we argue that there are some broader lessons emerging from our study that could support researchers in charge of designing these systems.

Despite the inherent similarities of the novel parallel system with the one previously used by workers in the café, all pilots stated that managing multiple avatars during their shift required a high cognitive demand. While pilots were able to successfully navigate interactions with customers at all stages, they reported being less able to focus on individual tasks as their attention needed to be split between multiple avatars performing different operations. This matches findings from literature focusing on media multitasking on attention [40]. Although, to a certain extent, the increase in cognitive demand and the resulting reduction of attention of single tasks are unavoidable, there are strategies that can be implemented to mitigate this. The review by Jeong & Hwang 2016 [40] shows that user control, task contiguity, and task relevance are all important moderators of multitasking effects that cause cognitive overload.

Based on our observations during the implementation study we hypothesize that the latter two moderators would remain equally valid regardless of the context. Despite not placing any constraints on how pilots could arrange the control interfaces of the different avatars on their screen, we notice that all of them chose to place the interface in close proximity to each other, to make it easier to switch attention between avatars through small gaze movements. Although it could be argued that for the two pilots (P1 and P2) who partially relied on eye gaze as a modality of input for mouse control, this was unavoidable the same behavior was observed across other pilots who accessed their computers using different modalities. Thus, facilitating attention shift between avatars with minimal effort remains an important design recommendation.

Task congruence was also an important mitigator for helping pilots to manage the cognitive load of parallel control, each avatar robot OriHime had specific functions that were well identified in the flow of operations that pilots were already familiar with. In a way, future implementation of parallel avatar systems should be considered in situations where it would make sense for users to take advantage of it to perform actions which are intrinsically connected. Moreover, task congruence can also be promoted by preserving similarities between the control interfaces of different avatars. While pilot needed training to be able to manage the simultaneous operations of avatars they were already proficient in managing the controls of a single avatar and they did not need to change the way they interacted with the interface. This also enabled them to maintain a consistent body language as they transition from one avatar robot OriHime to another.

Considerations around user control are slightly more complex and require careful examination of the context of a parallel control system and the purpose for which it is used. From a sensory point of view it is essential to give pilots complete control over the transition between different avatars. This includes key features such as selecting which cameras, speakers and microphones are active so that pilot can choose which avatar is currently listening to customers, or which one is looking around to navigate between tables. However, movement control might be limited to pre-defined actions to simplify the cognitive load of the pilot and reduce the complexity of the UI. Ultimately, we argue that a good tenet to follow in the design of parallel system is to give users as much control on each avatar as they need for the intended application, but no more that is needed to facilitate management of multiple avatars simultaneously, in line with the idea of smart systems that can provide assistance as required [35, 99].

### 6.3 More competent parallel pilots

One of the key aspects that played a significant role in the successful development and implementation of the parallel avatar systems is the expertise of the pilots who took active part in the project. Despite initial concerns about the complexity of the system, after the implementation all pilots expressed the opinion that the system could be easily integrated into the regular operations of the café. Although the operations training helped pilots to practice managing multiple avatars and feel more confident about their abilities, pilots were only able to learn how to smoothly manage complex operations quickly, because of the specific competencies that they already possessed as a result of their previous experience with telework in the café and beyond.

The idea that everyday technology practices enable disabled individuals to develop specific competencies that can be applied to a variety of situations has been formulated by several researchers.Reyes-Cruz et al [65] documented the ways in which visually impaired individuals navigate daily challenges and interact with technologies and illustrated how these led them to acquire skills in a variety of domains including, auditory, spatial, tactile as well social. These skills, it should be stressed, are not superhuman abilities that all individuals with a particular type of impairment magically acquire as a result of their disability (in the way it is often depicted by stereotypical super-crip portrayals [39]), but the result of everyday interactions that generate context specific epistemic knowledge [7].

In a similar fashion, pilots had developed specific competencies that boosted their ability to quickly master the parallel management of multiple avatar bodies. Being proficient in using avatar robots meant that pilots were comfortable experimenting with them, appropriating gestures and developing unique body languages, making the interactions with customers much more personal. Experience navigating the café using the avatar robot OriHime-D meant that pilot had a strong sense of special awareness, which has been reported as a common challenge for users of telepresence devices [33].

Pilots' competences are not just limited to avatar control alone, but encompass other key aspects of parallel interactions. Previous experience communicating customers' orders to the kitchen while continuing to entertain them meant that pilots had already developed effective media multitasking strategies that could be leveraged for parallel avatar control. Finally, strong social skills that helped pilots deliver high level customer service and provide memorable experiences to customers, were also employed to create playful interactions that could surprise and amaze, while also conveying the capabilities of the new parallel system.

These examples showcase how the unique blend of skills that café pilots possessed as a result of their practice of remote telework made them more apt to learn how to control and take advantage of parallel avatar bodies. Expanding the use of telepresence and robotic avatars to a greater variety of work environments might help businesses to access a skilled workforce that is currently excluded from many forms of employment due to the presence of accessibility barriers.

### 6.4 Accounting for context

The success of the development and implementation of the parallel system is, of course at least in part attributable to a series of factors linked to the context of the café itself and, to a certain extent, Japan more broadly. Firstly, the technical and human infrastructure of the Avatar Robot Cafe DAWN ver.$\beta$ minimised many of the issues that could have slowed or halted the progress of our study. Problems due to lack of connectivity or presence of environmental barriers that hinder the use of robotic avatars, which have been commonly reported in research [1, 70], did not affect the work of pilots in our study as the café represented an ideal setting for the parallel system. Secondly, Japan has been described as a moderately polychronic culture meaning that workers, disabled and non, might be more interested in interacting with systems that support multitasking [63]. Finally, previous research has shown that Japanese people are relatively open to the use of robots in various aspects of life, albeit not as much as other countries like the US [57, 58]. Establishing businesses like the Avatar Robot Cafe DAWN ver.$\beta$ or trialling systems for parallel avatar control could be less viable in countries were the general population is less interested or more resisting towards the use of avatar robots. At the same time, face-to-face interaction is extremely important in Japanese work culture, a circumstance that has hindered the mainstreaming of telework during the pandemic [18]. Countries where attitudes towards telework are more positive could see further opportunities for such systems.

## 7 CONCLUSIONS

Robotic avatars can be leveraged to enable disabled people to access work in the hospitality industry which is traditionally inaccessible to those who experience severe mobility limitations. However, constraints about avatars functionality and customer service operations management can limit the agency of disabled workers. In this paper we present the development and implementation of the first parallel system that allows disabled workers to control multiple robotic avatars simultaneously, increasing their agency and improving customer experiences. By conducting observations, surveys, and semi-structured interviews we explore the experiences of both workers and customers conceptualizing three different themes: Layers of interactions between people and avatars; Challenges, skills, and pride; Purpose and Play. Through reflections and comparison with relevant work we propose a series of design implications for future telepresence and avatar mediated interactions: 1) A different form of embodiment; 2) Balancing the need for more functionality with cognitive demand; 3) More competent parallel pilots.

## ACKNOWLEDGMENTS

We gratefully acknowledge the work of members of OriHime pilots who participated in this study. This work was supported by JST Moonshot R&D Program "Cybernetic being" Project (Grant number JPMJMS2013).